\documentstyle[12pt]{article} 

\newcommand{\be}{\begin{equation}}
\newcommand{\ee}{\end{equation}}
\newcommand{\bea}{\begin{eqnarray}}
\newcommand{\eea}{\end{eqnarray}}
\newcommand{\bean}{\begin{eqnarray*}}
\newcommand{\eean}{\end{eqnarray*}}

\def\thefiglist#1{\section*{Figure Captions\markboth
{FIGURE CAPTIONS} {FIGURE CAPTIONS}}\list
{Figure \arabic{enumi}.}
{\settowidth\labelwidth{Figure #1.}\leftmargin\labelwidth
\advance\leftmargin\labelsep
\usecounter{enumi}}
\def\newblock{\hskip .11em plus .33em minus -.07em}
\sloppy}

\makeatletter
\def\thebibliography#1{\section*{References\markboth
{REFERENCES}{REFERENCES}}\list
{\arabic{enumi}.}{\settowidth\labelwidth{#1.}\leftmargin\labelwidth
\advance\leftmargin\labelsep
\usecounter{enumi}\parsep \z@ \itemsep \z@ plus2pt}
\def\newblock{\hskip .11em plus .33em minus -.07em}
\sloppy
\sfcode`\.=1000\relax}

\makeatother
\newcommand{\gapproxeq}{\lower
.7ex\hbox{$\;\stackrel{\textstyle >}{\sim}\;$}}
\newcommand{\lapproxeq}{\lower
.7ex\hbox{$\;\stackrel{\textstyle <}{\sim}\;$}}

\begin{document}
\begin{titlepage}
\vskip 2cm
\begin{center}
{\Large\bf  Nucleon Spin: Summary}
\vskip 2cm
 {\bf \large F E Close\\
Rutherford Appleton Laboratory,\\
Chilton, Didcot, Oxon,
OX11 OQX, England.}\\
\end{center}
\begin{abstract}
This talk summarises the discussions during the conference on the spin
structure of the nucleon  held at Erice; July 1995.
The summary focuses on where
we have come, where we are now, and the emerging questions that direct  where
we go
next in the quest to understand the nucleon spin.
\end{abstract}
\end{titlepage}

We have spent a stimulating week discussing the status of the ``nucleon spin
puzzle".
At least we are agreed that neither is there nor, apparently, was there
ever,
any spin ``crisis".  We are also agreed that this field has been, and continues
to be, rich in opportunity and the unexpected.  I have always believed that the
essential
clues are implicit in an apparent paradox that shows up when one looks
at
the data in two complementary ways.  On the one hand, experiment measures
directly the polarisation asymmetry $A_1(x, Q^2)$.  World data
on both proton and neutron are in remarkable agreement with the ``pre-
historic" {\bf pre}-dictions
of
quark
models refs (1,2,3), at least for $x \gapproxeq 0.1$.  This would appear to
``confirm"
the
quark spins as primarily responsible for generating the nucleon spin (fig 1).

However, when we construct $g_1^p (x, Q^2) \equiv A^p_1(x, Q^2) F^p_1(x,
Q^2)$
and
integrate it, it falls short of the value expected by the Ellis-Jaffe sum rule
\cite{gourd}; typically for
SMC in 1994 at $Q^2= 10 GeV^2$
$$
I^p_{expt} = 0.136 \pm 0.011 \pm 0.011 ; I^p_{theory} = 0.176 \pm 0.006.
$$
The
discrepancy is some $2\frac{1}{2}\sigma$; the net quenching of inferred spin
$\Delta q$ is
roughly $\Delta q\simeq 9 (I^p_{theory}-I^p_{expt}$) and hence a shortfall of
0.04 in the integral magnifies
into
a quenching  of $\Delta q$ by some 40\%.  It is this dramatic shortfall that
has excited so
much
interest.  How is it that data presented one way (the $A_1(x, Q^2) $) appear
to
agree
so well with theory whereas the $\int dx g_1^p(x, Q^2) $ appears to give a
rather different message?

This meeting has sharpened understanding of this and helped to focus on the
leading current questions.

First,
construction of $g^p_1(x\rightarrow 0)$ is a particularly delicate issue, as
has
repeatedly been stressed.  Here are some issues that need study.\\
(i)  Experiment measures $A(x, Q^2) $.  This is $Q^2$ independent to good
accuracy
at $x\gapproxeq 3\times 10^{-2}$ and is {\bf assumed} to remain so even for $x<
10^{-2}$.  QCD evolution suggests this is not in general true, and the $Q^2$
dependence is a major issue.\\
(ii)  The above yields $A(x)\simeq$ constant as $x\rightarrow 0$\\ (iii) The
constant
$A(x)$ is then multiplied by $F_1 (x,Q^2)$ that rises as $x\rightarrow
0$.
To be pedantic, we know that $F_2 (x,Q^2)$  from NMC and HERA show a
marked
rise; more information on unpolarised structure, in particular on $R
(x,Q^2)\equiv
\sigma_L/\sigma_T$ and on the normalisation of NMC and HERA data sets
may
be needed before we can be entirely certain that there are no subtle errors
creeping in
here.

But one certain conclusion is that at $x<0.1$ we cannot lose.  Either
$g^p_1(x\rightarrow 0)$ behaves smoothly as has been assumed with the result
that
$\Delta q$
is quenched (hence theoretically challenging) or if $\Delta q$ is not
quenched there will be interesting behaviour to be measured at small $x$.

In this summary I shall start at large $x$ where we know what is going on but
don't
understand why.

It
is remarkable that the $x$-dependences of the polarisation asymmetries in the
valence region $A_1(x>0.2$) confirm the quark model predictions
\cite{kuti,fec74}
for  proton neutron and deuteron systems.  It is worth remembering that these
predictions were based initially on the assumption that the Pauli principle
correlates
the
spins and flavours of the valence quarks as in the familiar case of constituent
quarks
in spectroscopy.  These initial predictions were then modified in light of
the
emerging unpolarised $F_2^n(x)/F_2^p(x)$ and theoretical ideas concerning the
relationship between constituent and current quarks \cite{fec74}.

The region $x\rightarrow 1$ probes the deep valence structure of the current
quarks.  An untested prediction \cite{fec74} is that $A^n(x\rightarrow 1)
\equiv
A^p(x\rightarrow 1)$.
When $x < 0.3, A^n (x< 0.3) <0$;  thus an issue is whether
$A^n$  becomes positive
when $x\gapproxeq 0.4$.  The data error bars are too large to tell, hence the
challenge is to measure if
$$
\bullet A^n(x \gapproxeq 0.4) >0
$$
This
would at least be a qualitative indicator that the neutron is ``readjusting" so
as to
catch the large positive asymmetry of the proton.  The next question concerns
the
magnitude
of $A^N(x\rightarrow 1) $.  If a single quark carries all the helicity in
this
limit \cite{fec73,fec78} then $A(x\rightarrow 1)\rightarrow 1$.  If its spin is
quenched in line with the 25\% quenching of $g_A/g_V$ then $A(x\rightarrow
1)\rightarrow 3/4$.  However if a single flavour dominates (as suggested by
$F^n_2/F^p_2 (x\rightarrow 1)$) but it retains the naive SU(6) values, one has
$A(x\rightarrow 1)\rightarrow 2/3$.

Thus for the proton a first test may be
$$
\bullet Is A^p(x\rightarrow 1)>2/3 \; ?
$$
Showing whether $A> 3/4$ will be more difficult.

Do not overlook that CEBAF may be able to study $A(Q^2, W^2)$ in this region
of
$x$ but with $W^2$, the invariant mass squared of the hadronic system, tending
towards
the resonance region.  It will be interesting to have predictions on the
$x$
and $Q^2$ dependence in this limit where the spin response of the nucleon may
be
probed in some detail.

To the extent
that the polarisation of valence quarks is canonical, at least insofar
as
the asymmetry is concerned, we should confidently expect predictions for {\bf
asymmetries} of polarised $\Lambda$ etc. to apply in the valence region.
In turn the question arises as to what is
quenching the valence quarks' contribution to the net polarisation.

In 1977 Sivers and I already noted that the evolution equations of QCD imply a
nontrivial
polarisation of the sea.  First, helicity conservation implies that a
polarised
valence quark will bremstrahlung a gluon that will itself be polarised
with
the same
polarisation as that of the initial quark.  Hence, since $\Delta q_v >0$
then
$\Delta G>0$ also, at least at $0(\alpha_s)$.  We found that
$$
\frac{\Delta G(x\rightarrow 1)}{G(x)}
\equiv \frac{G^+(x)-G^-(x)}{G^+(x)+G^-(x)}
\simeq \frac{1-(1-x)^2}{1+(1-2)^2}
$$
which has been discovered independently more recently \cite{brodsky94}.
Experiment  E704 at
Fermilab \cite{E704} suggests that $\Delta G(x<0.3)$ is small: this may be
compatible
with the above since the gluon asymmetry is small there.  Measurement of
$\Delta
G(x>0.3)$ may be critical.

The next
stage in our paper was to study the implication for $\Delta q_{sea}(x)$.
A
polarised evolution $G^\uparrow (x) \rightarrow q\bar{q}$ indeed gives no net
helicity
in the sea but it does yield a local non-zero effect.  The gluon gives in
general $q(x_1)\bar{q}(x-x_1)$ where $0\leq x_1\leq x$.  The hard tail
$x_1\rightarrow x$ has $\Delta q=\Delta \bar{q} >0$ while the soft region
$x_1\rightarrow 0$ compensates with $\Delta q = \Delta \bar{q} <0$.  This
contrasts
with non-perturbative
effects such as a sea driven by $J^p=0^-$ meson clouds, for
which $\Delta \bar{q} <0$ (e.g. Isgur here \cite{isgur}).

Hence the challenge is to test whether, for the hard component at least,
\bea
\bullet& \Delta \bar{q} (x\; large) > 0\nonumber \\
\bullet &\Delta \bar{q} (x) <0\nonumber
\eea

The
sharpest probe, in theory, is to tag fast $K^-$ in the current fragmentation
region.  The idea \cite{milner} exploits the fact that $K^-(s\bar{u})$ contains
members
of the initial proton's sea and so, to the extent that the leading hadron
in a
jet contains the quark (or antiquark) that was struck by the current probe, the
$K^-$
is a
direct tag for the sea.  A polarisation asymmetry for the leading $K^-$ will
translate
into a polarised sea for the proton.  These questions are beginning to be
answered by SMC and will be  a major component of the HERMES programme.

Finally in this study of how the proton's spin is decomposed we have the ``sum
rule"
$$
\frac{1}{2} = \sum \Delta q + \Delta G + L_z
$$
where $L_z$ is the ``orbital angular momentum" of the constituents.
Discussions here show that there
 is  some confusion as to what ``$L_z$" means.  An example is given by QCD
evolution where $G^\uparrow (q\bar{q})_{\lambda =0}$ with $(\lambda =0)$
denoting
the net helicity of the $q\bar{q}$.  The gluon-$q\bar{q}$ vertex contains
an
$e^{i\phi}$,
where $\phi$ is the azimuthal angle of the $q\bar{q}$ plane relative
to
the gluon helicity and $<L_z> \simeq i\frac{d}{d \phi}$ represents the
transmutation of gluon helicity into the orbital angular momentum of the
$q\bar{q}$  (see e.g. ref \cite{rat}).  So in some sense $<L_z>$ measures the
number
of polarised gluons that has transmogrified into $q\bar{q}$.

Experimentalists are encouraged to seek $\phi$ dependence of the hadron
production \cite{Lz}; the theoretical and practical question is then how to
disentangle how
much of this is background from resonance decay or from quark-hadron
fragmentation.

\subsection*{Sum Rule Sensitivity and F/D}

Differences between $I^p_{exp}$ and $I^p_{theory}$ are magnified ninefold
when
interpreted as a quenching of $\Delta q$: consequently any apparently minor
adjustment
to the left hand side $(I^p_{exp}$) or right hand side (F/D)  of the EJ
sum rule can have an
order of magnitude impact on  the inferred magnitude of $\Delta q$.

One unresolved question is the interpretation of F/D when SU(3) is broken.
Experimental data on hyperon decays may still have something to offer.  For
example
$$
\frac{g_A}{g_V} (n\rightarrow p) =F+D\equiv \frac{g_A}{g_V}
(\Xi\rightarrow
\Sigma)
$$
so it
will be interesting to see if this equality is preserved when $\beta$-decay
occurs in the presence of spectator
strange quarks.  Secondly, for the case of strangeness
changing decays in the hadronic axial current
$$
A_\mu = g_A \gamma_\mu \gamma_5 - g_2
\frac{\iota\sigma_{\mu\nu}q^\nu \gamma_5}{m_i+m_j}
$$
it has been assumed that $g_2=0$.  While this is assured in the limit $m_i=m_j$
(such as
$n\rightarrow p)$ it is not necessarily so for $\Delta S=1$.  The Hsueh et
al
analysis of $\Sigma n$ made a fit \cite{hsu} allowing for $g_2 \neq 0$ and
found a
rather different value for $g_A$ than that taken from the Particle Data Group
\cite{pdg} and used in the extraction of F/D = $0.575\pm 0.016$ \cite{rgr}.  In
addition to these uncertainties there are systematic uncertainties due to phase
space
and form factors \cite{phil}.

Further
precision studies of hyperon decays may be warranted if the quantitative
precision on $\Delta q$ becomes an important issue.  For example, if $\Delta q$
is
quenched
it will be of interest to determine whether $\Delta q \simeq 0.3$ which
may be in the region of ``$\Delta G$ and the anomaly" \cite{mue} or whether
$\Delta q\rightarrow 0$ as in Skyrmion models \cite{ellis}.

My opinion is that one should continue  to use F/D $\simeq$ 0.58 and 3F-D
$\simeq$ 0.6
until proven wrong to do so.

\subsection*{$x\rightarrow 0$ Questions}

Historically extrapolation has used Regge with an $a_1$ trajectory whose
intercept
$\alpha (a_1)\simeq 0$.  However, Roberts and I  \cite{rgr2} originally noted
that diffractive behaviour in
spin dependence is a poorly understood area and that an $(x\log^2 x)^{-1}$
behaviour is allowed within the general Regge analyses.  So the first
question is, for Regge

$\bullet$  What extrapolation should one use?

$\bullet$  Up to what value of $Q^2$ is Regge legitimate?

Complementary to this is a renewed interest in the $x\rightarrow 0$ evolution
of
$g^p_1(x,Q^2)$ and the rise \cite{rgr2,ball}

$$
\bullet g_1 \sim \exp\sqrt{\ln 1/x}
$$
Some recent analyses suggest that there may be a rise even in the non-singlet
contribution \cite{ryskin}.  The questions here include

$\bullet$ At how small a value of $x$ do such ideas apply?

Kuti \cite{kuti95} has reanalysed the Regge theory and confirms the ``in
principle"
presence of $(x\ln^2 x)^{-1}$ but finds that its coefficient vanishes in the
Reggeon
calculus.  To settle the question of Regge behaviour empirically we need
to fix $Q^2$
at a small value common to SMC as $x\rightarrow 0$ and SLAC at $x \simeq
0.1$
say, and establish the energy dependence at fixed $Q^2$.

As to
whether/when Regge applies it is important to recall the historical origins
of its
application to deep inelastic.  When scaling was first observed, Regge theory
was a
leading idea.  It was noted that one could force a marriage between the two if
the
Regge residue had a magic behaviour.

$$
F_1(x,Q^2) \sim \beta(Q^2) \nu^\alpha \rightarrow x^\alpha \; if \; \beta
(Q^2)
\sim (Q^2)^{-\alpha}
$$
To my knowledge such a behaviour has not been derived from Reggeon field
theory!  This may be a question of interest to some, but suppose instead that
$\beta(Q^2) \sim (Q^2)^{-(\alpha+1)}$ in which case Regge applies as
$Q^2\rightarrow 0$ but is rapidly overtaken by QCD evolution (to which it
might
have no immediate relation).  Thus the energy dependence of low $Q^2$ data
may
be quite different to that at high $Q^2$:  HERA data on $F_2(x,Q^2)$ may give
insight into this general question.

Until these questions are better understood we may gauge the ``theoretical"
systematic errors on $\Delta q$ by extrapolating with
\bea
\bullet & x^\alpha\nonumber \\
\bullet& (x\ln^2 x)^{-1} \nonumber \\
\bullet &\ln  x\nonumber \\
\bullet  &\exp\sqrt{\ln 1/x} \nonumber
\eea

The resulting range on $\Delta q$ may be larger than other errors and this at
least
would highlight what are the most important issues.  Even so, they are unlikely
to
raise $\Delta q$ to the naive valence of $\simeq 0.6$ when one treats both
proton
and neutron (deuteron) target data simultaneously (see later).

\subsection*{The Erice Statement}

It is
agreed that one plot $xg_1(x)$ against $\log x$ or one plots $g_1(x)$ against
$x$
when attempting to visualise the measurement of sum rules

$$
I \sim \int dx g_1(x) = \int d (\ln x) xg_1 (x).
$$
Plotting $g_1(x)$ against $\log x$ is to be used only for making propaganda and
will
be recognised as such.

I shall now show $g^p_1$ and $g^n_1$ plotted against $\log x$!  This expands
the $x
< 10^{-1}$ region and highlights a marked difference between the proton and
neutron (fig 2).  This is an interesting area demanding further study.  I will
motivate
this
by recalling why the deuteron target has interest.

The generic sum rules have the structure for target A.

$$
I^A = a I_3 + bI_8 + c \Delta q
$$

Using F/D $\simeq$ 0.58 to relate $I_8$ to $(g_A/g_V$) and including QCD
corrections we may write
$$
I^A=A(\frac{g_A}{g_V}) + B\Delta  q
$$
Very approximately (ignoring $0(\alpha_s$))
\bea
I^p \simeq \frac{1}{10} (\frac{g_A}{g_V}) + \frac{1}{9} \Delta q\nonumber\\
I^n \simeq -\frac{1}{15} (\frac{g_A}{g_V}) + \frac{1}{9} \Delta q\nonumber
\eea
so that
\bea
I^{p-n} =\frac{1}{6} (\frac{g_A}{g_V}) \nonumber\\
\frac{1}{2} I^{p+n} =\frac{1}{60} (\frac{g_A}{g_V}) + \frac{1}{9} \Delta
q\nonumber
\eea
Thus the $p-n$ difference is the Bjorken sum rule for which $\Delta q$ vanishes
and $p+n$ is the best
for emphasising   $\Delta q$.  In 1988 we expected  \cite{rgr}
that for the
deuteron the $g_1^d(x>x_c)>0$ where $x_c \lapproxeq 0.1$.  The predicted (and
now empirical) $g^n_1<0$ as $x\rightarrow 0$ gave the possibility that
$g_1^d(x<x_c)<0$.  If so one would have an upper limit on  $\Delta q$  without
any
need to worry about extrapolating to $x=0$.
$$
\int^1_{x_c} dx g_1^d (x,Q^2) \leq \int^1_0 dx g^d_1 (x,Q^2)
$$
Using the present data gives  $\Delta q \lapproxeq 0.25$.

However there is a catch.  We assumed that $g_1^d(x\rightarrow 0$) does not
oscillate, i.e.
does not become positive at even smaller $x$.  In this context the
SMC
datum $x=5\times 10^{-3}$ is
tantalising (fig 3).  The challenge will be to reduce
the errors
on this datum to see whether $g_1^d (x\simeq 5\times 10^{-3}) >0$.  If
$g_1^p(x\rightarrow
0$) is indeed rising due to a singlet (diffractive) dominant contribution, then
$g^d_1$ will have to become positive too.

The fact that  $g_1^p \neq g^n_1$ for $x \gapproxeq 5\times 10^{-3}$ shows that
there is substantial non-singlet, non-diffractive, contribution still present.
Indeed
one may note that to a reasonable approximation that
$$
g^n_1 \simeq - g^p_1
$$
in a substantial region.
Kuti and Roberts \cite{kutirgr} have even noted that an
extreme simultaneous fit would allow $a_1$ exchange with $\alpha \simeq$ 0.3.
This large intercept is required to accommodate the rise in $g_1^p(x\rightarrow
0$)
and would give $\int^0_{0.003} g^p_1 (x) \simeq 0.025$ with consequent extra
contribution to $\Delta q \simeq 0.2$
and elevating the total $\Delta q$ towards
the naive quark model value.  This would be a dramatic conclusion if
confirmed.  Problems include why $a_1$ exchange totally dominates $f_1$ (in
unpolarised Compton scattering the analogous $a_2:f_2$ is only in ratio 1:5).

\subsection*{$g_2(x)$}

The Burkhardt-Cottingham sum rule  $\int dx g_2 (x,Q^2) =0$ needs to be tested
but how much effort is
needed to confirm zero is zero?  The main challenge will be to measure or limit
$\bar{g}^2(x,Q^2)$ where
 $$
\bar{g}_2(x,Q^2) \equiv g_2(x,Q^2) - [\int^1_x  g_1(y,Q^2) d \ln y -g_1(x,Q^2)]
$$
This is a unique and {\bf direct} measurement of twist 3 contributions: other
tests of
higher twist tend to rely on model dependent fits to data.

\subsection*{$b_1(x)$}

For $J=1$ targets this can be non-zero.  However, before investing too much
experimental effort on the deuteron bear in mind that, to the extent that the
deuteron is made of two independent spin 1/2 components, $b_1(x)\rightarrow
0$.
There is a parton model sum rule \cite{kum}
$$
\int dx b_1(x) =
\lim_{t\rightarrow 0} t F_Q(t) + \frac{1}{9} \delta Q\rightarrow
\frac{1}{9} \delta Q
$$
where $F_Q(t)$ denotes the quadrupole moment of the target and $\delta Q$ is
the
quadrupole polarisation of the sea $\equiv  q^0_\uparrow  -\frac{1}{2}
(q_\uparrow
+ q_\downarrow)^1$ where the superscripts denote the target $\hat{z}$
polarisation and $q$ denotes $q+\bar{q}$ number.  There are further spin
dependent
structure functions, such as $h_1(x)$ \cite{jaffe} which may be probed at
polarised
RHIC or by the HMC collaboration at CERN within a few years.

\subsection*{The $Q^2\rightarrow 0$ Polarisation Asymmetries}

To my knowledge it was Gilman in 1971 \cite{gilman} who first questioned how
the
Bjorken sum rule  and predictions of a large {\bf positive}  $A_1(x)$ for the
proton
would match with the requirements of the DHG sum rule \cite{dhg} that
$\langle A \rangle <0$
for $Q^2=0$.  We discovered that the quark model predicted that a rapid change
from $A<0$ to $A>0$ would occur in the resonance region, a phenomenon that
was
subsequently confirmed by experiments in 1973 for the prominent $N^*(1520)$
and
$N^*(1690)$ \cite{cg72}.  The change in sign occurs for $Q^2<0.5 $ GeV$^2$ for
the
latter and possibly even by $Q^2$=0.3 GeV$^2$ for the former.  In the
$\Delta$(1230)
we expect that the
resonance excitation cross section drops very fast with $Q^2$,
revealing the $\pi N\; S$-wave background (with $A>0$).

The only  direct measurements of total asymmetries through the $N^*$ region
show $A(Q^2\geq $0.5 GeV$^2)>0$.  However, even this is at too large a $Q^2$
for
our purposes!
It is necessary to understand how the asymmetry changes sign as a
function of
$Q^2$  and $W^2$ .  If it changes sign at $Q^2$=0.3 GeV$^2$ for all
$W$,
then it will be
irrelevant as a higher twist phenomenon that potentially affects
the
interpretation of the
Ellis-Jaffe sum rules. \cite{ioffe}.  However, if the sign
change
occurs at (approximately) constant $Q^2/W^2$ , then the region $x\lapproxeq
0.2$
could exhibit interesting non-perturbative $Q^2$  dependence at values of
$Q^2\gapproxeq$ 1 GeV$^2$

The programme at CEBAF should provide some insights.  In addition to the
above
it promises to probe the spin dependence of transitions $p\rightarrow N^*$ in
detail, thereby revealing the electric and magnetic response as a function of
quantum numbers (angular momentum or multipoles).  This will be a fine
detail
probe of the constituent valence nucleon structure whose relation with the deep
response of valence current quarks may reveal the profound and poorly
understood
transformation from constituent to current quarks.

\subsection*{HERA and Gluemorons}

The rapidity gap events at HERA may be interpreted on rather general grounds
as
due to the proton offering up a colour singlet non-baryonic system whose
misconstructure is then probed by the virtual photon.  This ``Pomeron" may be
made primarily of
glue or of quarks.  Let's refer to these extreme possibilities as
``Gluemeron" or ``Quarkball".  The question is: how on general grounds can one
distinguish between these two broad classes?

The answer \cite{forshaw} is to focus not on the $x$-dependence of the object's
$F_1(x,Q^2$) but its $Q^2$ dependence, specifically

$$
\frac{d p}{dQ^2}  \equiv \frac{d }{dQ^2} \int^1_0 dx F_2 (x,Q^2)
$$
For a quarkball
(think of the familiar proton as an example) this falls gradually to
an
asymptotic value, whereas for a glueball or a gluemoron it should rise rather
rapidly
to this asymptopia.

The essential reason is that quarks shed momentum by gluon bremstrahlung
whereas gluons feed momentum into $q\bar{q}$. Hence in regions of $x$ where
quarks dominate, the
$\frac{d F_2(x)}{dQ^2} <0$ with increasing $Q^2$; by contrast
$\frac{d F_2(x)}{dQ^2} >0$ in regions of gluon dominance.

Gluon dominance is anticipated as $x\rightarrow 0$ for all systems and hence
$\frac{d F_2(x\rightarrow 0)}{dQ^2} >0$ in general.  For a quarkball this is
compensated at large $x$ where valence quarks dominate with the consequence
that
$\frac{d p}{dQ^2} <0 $ overall.  For a gluemeron however, the ``valence" gluons
cause $\frac{d F_2(x>0.3)}{dQ^2} >0$  which is quite opposite to that of a
quarkball.

Eventually data at HERA may quantify the $\frac{d p}{dQ^2}  $.  However it is
already apparent at $x\simeq 0.6$ that $\frac{d F_2(x=0.6)}{dQ^2} {\slash <}0 $
and
is
probably positive.
Thus it seems likely that there is strong indication that the
Pomeron is a gluemeron, independent of particular model dependent
assumptions
as to $x$ dependence.

Having established the gluemeron, we may look forward 25 years to polarised
HERA.  In addition to measuring $g^p_1(x<10^{-4}$) it may also probe the spin
structure of polarised gluemerons.

If the
gluemeron has $J=0$ there will be no spin asymmetry.  If it has $J\neq 0$
we
will
need to know the probability that it is offered up with helicity parallel or
antiparallel to the probe.  Suppose we have solved this and are doing polarised
deep
inelastic scattering from a polarised gluemeron as the latter evolves into
$q\bar{q}$
with $Q^2$.

Will the asymmetry be positive  (the $q\bar{q}$ remember the gluemeron
polarisation),
zero (the gluon helicity turns into $L_z$ in evolution) or negative
(the anomaly reads $-\alpha \Delta G$).  I asked several theorists to choose
among
these three and there was a roughly equal split between four answers.

It is 25
years since theorists first predicted the valence polarisation for quarkballs
and
it may be 25 years before we know the answer for gluemerons.  Vote now.

\subsection*{A Moral for Fundamental Curiosity Driven Research}

Many politicians believe that you have to be able to see the endgame if any
research
is to be worthwhile.  We all know counter-examples from Maxwell, Faraday, X-
rays
etc.  I added one to my list this week in the opening address by Vernon Hughes
who
nearly 40 years ago ``was stimulated  by parity violation" to make polarised
electrons
``with no obvious use".  Who would fund such blue skies today?  ``Then in 1968
quarks became real!"  The rest, as they say, is history.

\begin{thefiglist}{5}
\item{ } Predictions for $A^{p,n}$ in the valence region compared with recent
data.  The curves (ref 1) correspond to $A^p_1=(\frac{19-6R}{15})\xi$,
$A^n_1=(\frac{2-3R}{5R})\xi$ with $R=F^n_1/F^p_1$ and $\xi$=1 (solid),
$\xi$=0.75 (dashed).  See ref 2 (ii) for more details
\item{ } $g^p_1$ and $g^n_1$ versus $\log x$
\item{ } $g^d_1$ versus $\log x$
\end{thefiglist}


\begin{thebibliography}{45}
\bibitem{kuti} J Kuti and V Weisskopf, Phys Rev {\bf\underline  D4} (1971) 3418
\bibitem{fec74} F E Close, Nucl Phys {\bf\underline B80} (1974) 269;\\ ``The
Nucleon Spin Crisis Bible" RAL-93-034, Proc of 6th ICTP Workshop, Trieste
\bibitem{fec73} F E Close, Phys Lett {\bf\underline 43B} (1973) 422
\bibitem{gourd} M Gourdin, Nucl Phys {\bf\bf\underline B38} (1972) 418;\\
J Ellis and R L Jaffe, Phys Rev {\bf\bf\underline D9} (1974)  1444
\bibitem{fec78} F E Close and D Sivers, Phys Rev Letters {\bf\underline 39}
(1977) 1116
\bibitem{E704}  E704 Collaboration, Phys lett {\bf\underline  B336} (1994) 269
\bibitem{brodsky94} S J Brodsky et al, Nucl Phys {\bf\underline B441} (1995)
197
\bibitem{isgur} N Isgur, these proceedings
\bibitem{milner} F E Close and R Milner, Phys Rev {\bf\underline D44} (1991)
3691
\bibitem{rat}P G Ratcliffe,  Phys Lett {\bf\underline B192} (1987) 180
\bibitem{Lz} M Ta-chung et al, Phys Rev {\bf\underline  D40} (1989)  769
\bibitem{hsu} S Hsueh et al., Phys Rev {\bf\underline D38} (1988) 2056
\bibitem{pdg} Particle Data Group, Phys Rev  {\bf\underline D50} (1994) 1173
\bibitem{rgr} F E Close and R G Roberts, Phys Rev Letters  {\bf\underline 60}
(1988) 1471; Phys Lett  {\bf\underline B316} (1993) 165
\bibitem{phil} P D Ratcliffe, Phys Lett  {\bf\underline  B242} (1990) 271
\bibitem{mue} A Mueller, these proceedings
\bibitem{ellis} J Ellis, these proceedings
\bibitem{rgr2} F E Close and R G Roberts, Phys Lett
{\bf\underline  B336} (1994) 257
\bibitem{ball} R Ball, these proceedings
\bibitem{ryskin} J Bartels, B Ermolaev and M Ryskin, DESY-95-124
\bibitem{kuti95} J Kuti, these proceedings
\bibitem{kutirgr} J Kuti, private communication; R G Roberts, private
communication
\bibitem{kum} F E Close and S Kumano, Phys Rev {\bf\underline  D42} (1990)
2377
\bibitem{jaffe} R Jaffe, these proceedings
\bibitem{gilman} F J Gilman, question to FEC, SLAC 1971
\bibitem{dhg} S D Drell and A C Hearn, Phys Rev Lett
{\bf\underline  16} (1966)
908; \\S B Gerasimov, Sov J Nuc Phys  {\bf\underline  2} (1966) 430
\bibitem{cg72} F E Close and F J Gilman, Phys Lett  {\bf\underline  38B} (1972)
541;\\
 F E Close, F J Gilman and I Karliner, Phys Rev
{\bf\underline  D6} (1972) 2533
\bibitem{ioffe} B Ioffe, these proceedings
\bibitem{forshaw} F E Close and J Forshaw,  ``Are Diffractive Events at HERA
due to a Gluemeron or a Quarkball?" RAL-95-046, hep-ph/9509247
\end{thebibliography}
\end{document}